\documentclass[pdflatex,sn-mathphys-num,iicol]{sn-jnl}

\usepackage{graphicx}%
\usepackage{multirow}%
\usepackage{amsmath,amssymb,amsfonts}%
\usepackage{amsthm}%
\usepackage{mathrsfs}%
\usepackage[title]{appendix}%
\usepackage{xcolor}%
\usepackage{textcomp}%
\usepackage{manyfoot}%
\usepackage{booktabs}%
\usepackage{algorithm}%
\usepackage{algorithmicx}%
\usepackage{algpseudocode}%
\usepackage{listings}%
\usepackage{geometry,lipsum,adjustbox,graphicx,float,gensymb,cancel,mathrsfs,url,caption,subcaption,cancel,amsfonts,textcomp,tikz,xcolor,wrapfig,,sidecap,pdfpages,fancyhdr,booktabs,multirow,capt-of,tabularx,changepage,transparent,etoolbox,titlesec,rotating,stfloats,pdflscape,ltablex,mdframed,tikzpagenodes,amsmath}

\theoremstyle{thmstyleone}%
\theoremstyle{thmstyletwo}%

\theoremstyle{thmstylethree}%

\begin{document}

\title[Article Title]{Field validation of GNSS-independent positioning enhancement using a wearable ultra-stable quantum magnetometer}


\author*[1]{\fnm{Stirling} \sur{Scholes}}\email{s.scholes@strath.ac.uk}
\author[1]{\fnm{Dominic} \sur{Hunter}}\email{d.hunter@strath.ac.uk}

\author[1]{\fnm{Courtney} \sur{Dyer}}\email{courtney.dyer@strath.ac.uk}
\author[1]{\fnm{Marcin} \sur{Mrozowski}}\email{marcin.mrozowski@strath.ac.uk}
\author[1]{\fnm{Allan} \sur{McWilliam}}
\author[1]{\fnm{Phoebe} \sur{Utting}}
\author[2]{\fnm{David} \sur{Burt}}
\author[1,3]{\fnm{Paul F.} \sur{Griffin}}\email{paul.griffin@strath.ac.uk}
\author[1,3]{\fnm{James} \sur{McGilligan}}\email{james.mcgilligan@strath.ac.uk}
\author[1,3]{\fnm{Erling} \sur{Riis}}\email{e.riis@strath.ac.uk}
\author[1,3]{\fnm{Stuart J.} \sur{Ingleby}}\email{stuart.ingleby@strath.ac.uk}

\affil[1]{\orgdiv{Department of Physics}, \orgname{SUPA, University of Strathclyde}, \orgaddress{ \city{Glasgow}, \postcode{G4 0NG}, \country{United Kingdom}}}

\affil[2]{ \orgname{Kelvin Nanotechnology, University of Glasgow}, \orgaddress{ \city{Glasgow}, \postcode{G12 8LS}, \country{United Kingdom}}}

\affil[3]{ \orgname{Quantrologee}, \orgaddress{ \city{Glasgow}, \postcode{G12 8QQ}, \country{United Kingdom}}}




\abstract{Increasing the resilience of positioning systems that currently rely on Global Navigation Satellite System (GNSS) signals can be achieved by incorporating stable and sensitive measurements of the permanent crustal anomalies in the Earth's magnetic field. We have realised this concept using an in-house-developed, wearable, Free-Induction-Decay Optically Pumped Magnetometer (FID-OPM) to carry out precise and stable measurements of the geomagnetic field in a walking trial. We present an end-to-end validation, including qualification of FID-OPM performance, alongside quantification of improvement in accuracy when data from this sensor is added to a dead-reckoning estimation of position. Using our wearable sensor system we achieve a Beckmann-distributed radial positioning error of 2.24~m over a route exceeding 500 m in length and spanning $\sim$360 s.}




\maketitle

\section{Introduction}
\label{sec:Introduction}

Over the last four decades, Positioning, Navigation and Timing (PNT) applications have become increasingly reliant on Global Navigation Satellite Systems (GNSS). Near-global availability, combined with relatively inexpensive and scalable receiver hardware, has resulted in ubiquitous reliance on GNSS for PNT use-cases ranging over many civil, industrial, and military applications~\cite{li2026}. However, GNSS has a number of vulnerabilities; with a ground-level signal strength typically below -125~dBm, GNSS signals are easily jammed, spoofed or interrupted~\cite{Ioannides2016,UKriskreg,Liu2021}; GNSS coverage is reduced at the Earth's poles, in built-up areas \cite{Isik2020}, and unavailable underground and underwater~\cite{Tomita2024}.\\ 
\begin{figure*}[b!]
\begin{center}
  \includegraphics[width=0.95\linewidth]{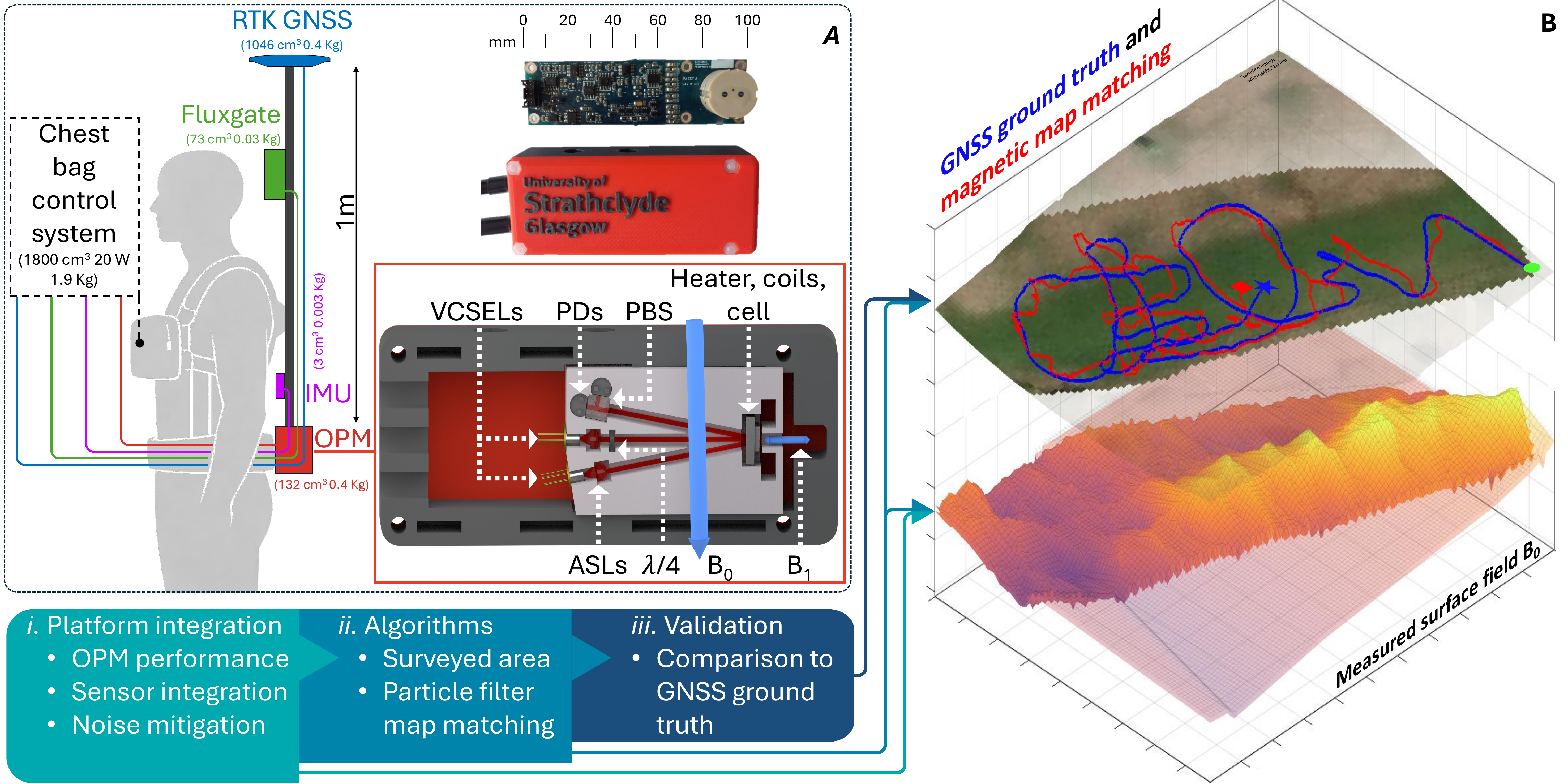}
  \caption{\textit{A} The Size Weight and Power (SWaP) characteristics of the harness-based wearable system. IMU: Inertial Measurement Unit, RTK GNSS: Real-Time Kinematic GNSS. The inset shows the functional components of the OPM. VCSELs: Vertical Cavity Surface Emitting Lasers, PDs: PhotoDiodes, PBS: Polarising Beam Splitter, ASLs: A-Spherical Lenses, $\lambda/4$: Quarter waveplate, $B_0$: Earth field, $B_1$: Enhanced optical pumping field. The scale shows a size comparison to a Consumer Off-The-Shelf (COTS) fluxgate. \textit{B} The blue boxes labelled \textit{i} through \textit{iii} show the processes that lead to the 3D visualisation on the right-hand side. The lower plane shows the magnetic map interpolated from the measured OPM survey (Fig.~\ref{fig:survey_map_no_interp}). The upper plane shows the surface topography, the ground-truth path (blue), and the magnetic map-matched path (red). The paths start at the green dot and end at the red-square and blue-star respectively. Regions outside of the surveyed area are made transparent. Full analysis of the methodology and results are given in the relevant sections.}
  \label{fig:setup}
\end{center}
\end{figure*}\\
For use-cases requiring resilient or ubiquitous positioning, GNSS can be augmented by additional positioning aids, which include reference to mapped signals of opportunity, such as magnetic, geodetic or gravitational topography, and reference to fixed ground-based transmitters, for example the enhanced LOng RAnge Navigation (eLORAN) system~\cite{Williams2013}. Better positional hold-over in the absence of GNSS signals can also be achieved through dead-reckoning, i.e., the integration of platform-mounted inertial sensing and timing references. Across all of these resilient PNT tools, new generations of quantum technologies offer enhanced capabilities; accurate quantum magnetometers~\cite{Schoenau2025,Kiehl2024,Wang2025,Bulatowicz2012} and gravimeters~\cite{Stray2022} offer calibration-free portable sensors, making map matching feasible over extended periods; portable optical clocks~\cite{Ducoing2025, Mistry_2026} extend the distribution of accurate phase coherence in beacon networks and underpin improved positioning by dead-reckoning; atom interferometers~\cite{Lee2022} and nuclear spin gyroscopes~\cite{Zhang2026} augment classical inertial sensor technologies with ultra-low scale factor drift and high bias stabilities.\\ 
\\
Magnetic map matching, i.e., positioning inference by the comparison of measured time series magnetic data from a moving platform to a fixed map of geomagnetic anomalies, interpolated from prior magnetic survey data, is a useful complement to a suite of resilient positioning tools~\cite{Canciani2016,Canciani2022,Jukic2024} and compares favourably to many other alternative positioning aids. Specifically, the crustal lithospheric magnetic anomaly field varies over geological timescales, and is effectively static over the duration of human journeys. Unlike visual positioning, magnetic anomalies are available globally, even over and under the oceans, and do not vary with lighting or weather. Radio and communication signals of opportunity can provide high levels of accuracy, but rely on infrastructure, which can be jammed or degraded. The large spatial extent of geomagnetic crustal anomalies makes them impossible to spoof or degrade (at scale) with anthropogenic magnetic sources. Measurement of these anomalies is effectively passive, and so can be carried out covertly without emission of detectable signals. Lastly, when compared with gravitational anomaly map matching~\cite{Lellouch_2025}, the contrast of magnetic anomaly signals is significantly higher and can be carried out with higher cadence. Typical crustal magnetic field anomalies are around $\sim$100~nT~\cite{Scholes2026}, and can be measured with pocket-sized devices of sensitivity $\sim$~60~fT/$\sqrt{\text{Hz}}$ and bandwidths $\sim$~5~kHz~\cite{Schoenau2025}. Crustal gravitational anomalies are around $\sim$10~mGal, and can be measured with portable devices having sensitivities of $\sim$~18~$\mu$Gal/$\sqrt{\text{Hz}}$ and bandwidth $\sim$~1~Hz~\cite{Prasad2022}.\\ 
\\
Magnetic crustal positioning could therefore be a practical and viable addition to a resilient navigation system, analogous to the manner in which GNSS is currently augmented by independent inertial and timing instrumentation~\cite{Muradoglu:2025blh}. However, to develop these systems there are a number of technical and integration challenges which must be addressed. The Earth's magnetic field is subject to variation due to the effects of space weather transients and diurnal fluctuations, and the ionosphere acts as a resonant waveguide for Ultra-Low Frequency (ULF) and Extra-Low Frequency (ELF) magnetic transients~\cite{Beggan2018,Bianchi_Meloni_2007}. For positioning, these are sources of spurious magnetic variation in the pT to nT range. As well as these atmospheric error sources the hardware and systems of the platform may give rise to other unwanted magnetic field variations. Further to these challenges, crustal anomaly reference data, while extant over virtually all of the Earth's surface~\cite{Maus2007,Beamish2011}, consists of survey measurements of widely varying density, survey date, instrumentation, and altitude. Therefore, the use of these reference data requires efficient interpolation methods to allow calculation of expected magnetic anomaly fields over possible positioning locations \cite{Kay_Dimitrakopoulos_2000,Wang2019}.\\
\\
In the context of these challenges, a successful demonstration of positional aiding using Optically Pumped Magnetometers (OPMs) must demonstrate several parallel capabilities. First, a precise and accurate sensor, integrated with its platform, with effective mitigation of platform-induced magnetic fields. Second, the application of suitable algorithms to interpolate accurate magnetic anomaly survey data and fuse magnetic and inertial sensor data into positional estimates. Third, the quantitative validation of the methodologies used i.e., trials must be conducted with an accurate ground truth and the additional benefit of the OPM integration determined.\\ 
\\
In this trial we have demonstrated these aforementioned parallel capabilities as shown by the progression of boxes \textit{i} through \textit{iii} in Fig.~\ref{fig:setup}. We report the sensitivity and accuracy of an OPM sensor as qualified in the laboratory against a fluxgate magnetometer of similar form factor and leverage the low Size Weight and Power (SWaP) characteristic of our sensor to develop a wearable system. As shown in Fig.~\ref{fig:setup}\textit{A}, the system integrates control and power systems, data acquisition, auxiliary sensors, and a Real-Time Kinematic (RTK) GNSS receiver. We report a ground-based magnetic survey in a rural location, implementing algorithms to build the magnetic map shown in Fig.~\ref{fig:setup}\textit{B}. We make selective use of GNSS, using it only for initial steps such as mapping, and post trial for results comparison. We demonstrate positioning enhancement with the application of a particle filter based map matching scheme. The map-matched path is shown by the red route ending at the red square on the upper layer on Fig.~\ref{fig:setup}\textit{B}. We quantify this improvement by comparison to a GNSS ground truth as shown by the blue path terminating at the blue star.
\section{Results}
\label{sec:Results}
\begin{figure}[t!]
\begin{center}
  \includegraphics[width=0.95\linewidth]{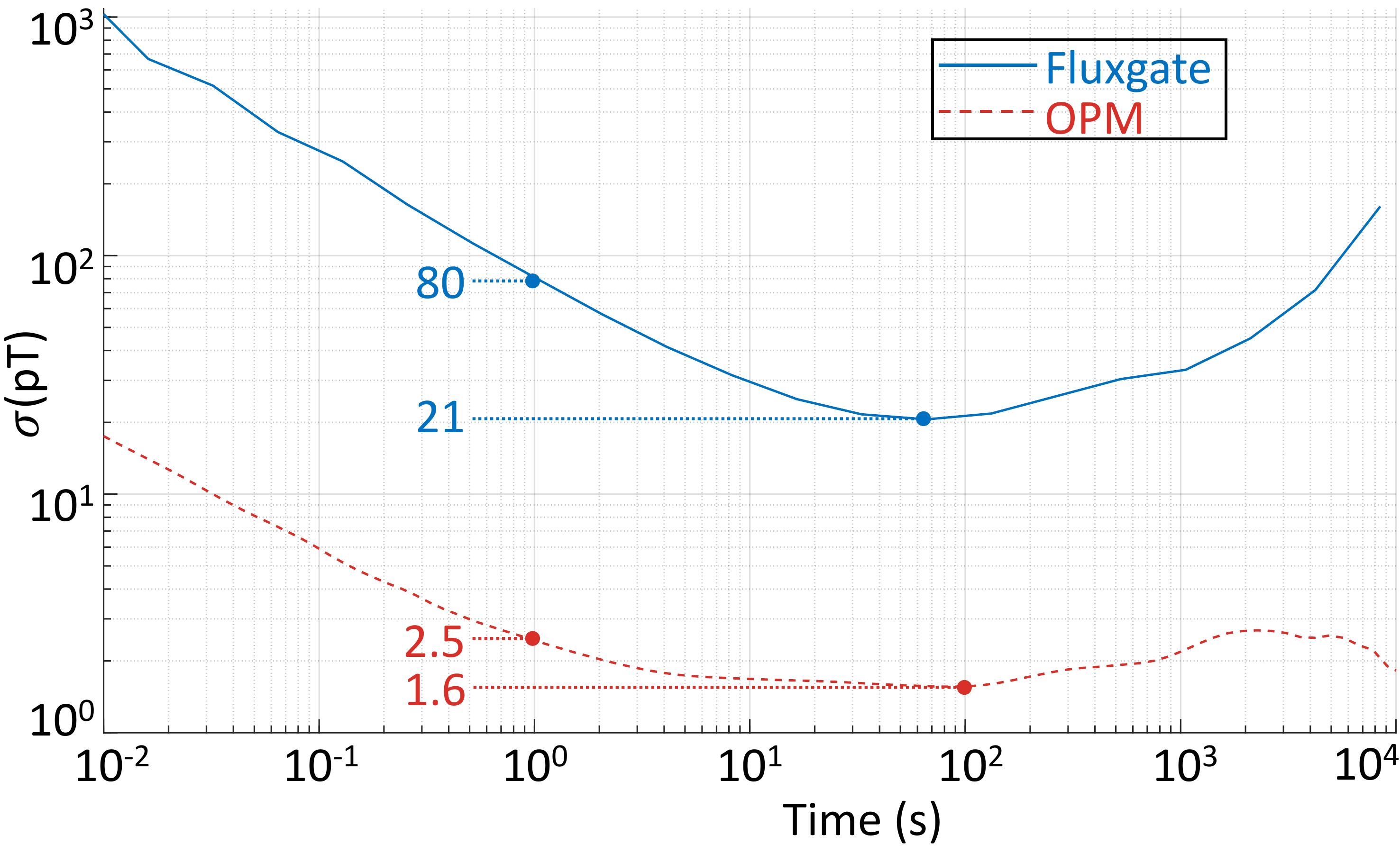}
  \caption{The measured Allan deviation over 10,000~s for the fluxgate (blue) and FID OPM (red). The markers show, the values at a time of 1~s, and the global minimum for each line.}
  \label{fig:a_dev}
\end{center}
\end{figure}
Our approach is based around a Free Induction Decay (FID) OPM developed at the University of Strathclyde as shown in Fig.~\ref{fig:setup}. Figure~\ref{fig:a_dev} characterises the stability of the OPM in comparison to the fluxgate used in the field trial (Bartington 901 and Waveshare high-precision PiHat analogue to digital converter). To characterise the sensors they were independently placed inside a 5 layer mu-metal shield in which a stable 10~$\mu T$ field was generated~\cite{mrozowski2023ultra}. The data was logged using the supporting electronics shown in Fig.~\ref{fig:setup}\textit{A}. Figure~\ref{fig:a_dev} shows that the OPM is able to achieve sensitivities more than an order of magnitude better than the fluxgate, $\sim$2 pT versus $\sim$ 80 pT at 1 s, while maintaining longer term stability.\\
\begin{figure}[b!]
\begin{center}
  \includegraphics[width=0.95\linewidth]{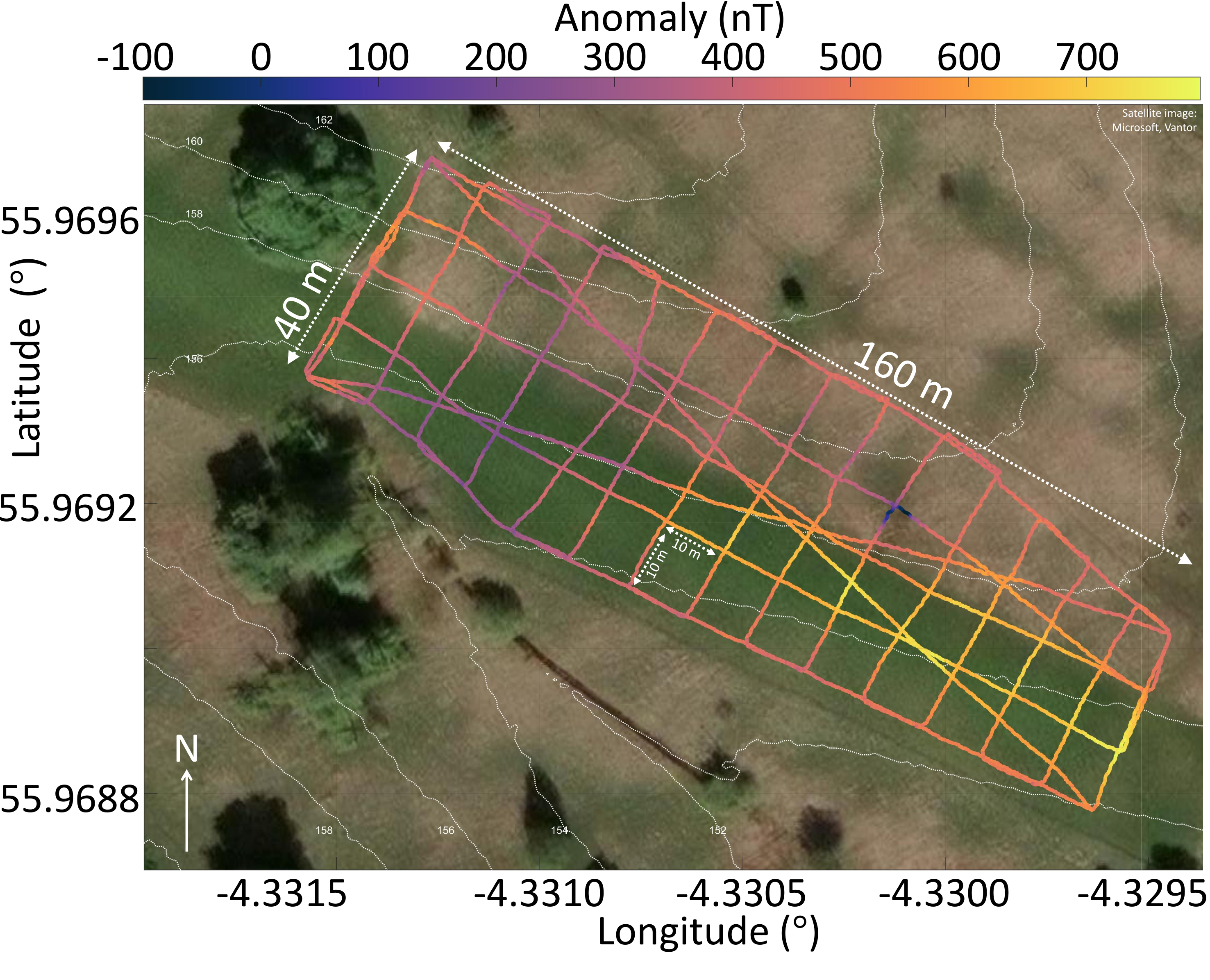}
  \caption{The surveyed data for the magnetic map. A raster path with 10 m line spacing was walked, immediately followed by two diagonal lines.  The thin white-dotted lines and accompanying numbers are the contour lines of the topography, illustrating the Northerly slope of the survey area.}
  \label{fig:survey_map_no_interp}
\end{center}
\end{figure}\\
Figure~\ref{fig:survey_map_no_interp} shows the surveyed data for the magnetic map gathered using the setup shown in Fig.~\ref{fig:setup}\textit{A}. An area of approximately 0.64 hectares (6400~m$^2$) was surveyed using a raster grid with 10-m spacing using the GNSS system to record the position of the magnetic samples. The average positional accuracy reported by the GNSS system with respect to the base station throughout the trial was 1.6~cm. Additionally, two diagonal `algorithm calibration lines' were also walked. The raw OPM data was processed and reduced to an anomaly value (Section~\ref{subsec:Data processing, calibration, and mapping} for details). Values from a local base station equipped with a fluxgate were compared against data from the British Geological Society's magnetic observatory at Eskdalemuir ($\sim$100 km from the test site) and no significant space weather deviations were noted during the survey period. Figure~\ref{fig:survey_map_no_interp} shows that even at relatively local scales, the measured field at the surface contains notable anomalies, with the field in our surveyed area spanning $\sim 800$ nT. The magnetic intensity shown in the 3D panel of Fig.~\ref{fig:setup}\textit{B} is interpolated from the samples shown in Fig.~\ref{fig:survey_map_no_interp} excluding the diagonal lines. The magnetic intensity shows a `magnetic ridge' along most of the Southern border the survey area, flattening to a plateau along the Northern perimeter, and a series of magnetic depressions in the North-Western corner. The abrupt magnetic anomaly near 55.9692~N -4.3300~W is a mild steel wire fence enclosing a sapling.\\   
\begin{figure*}[t!]
\begin{center}
  \includegraphics[width=0.95\linewidth]{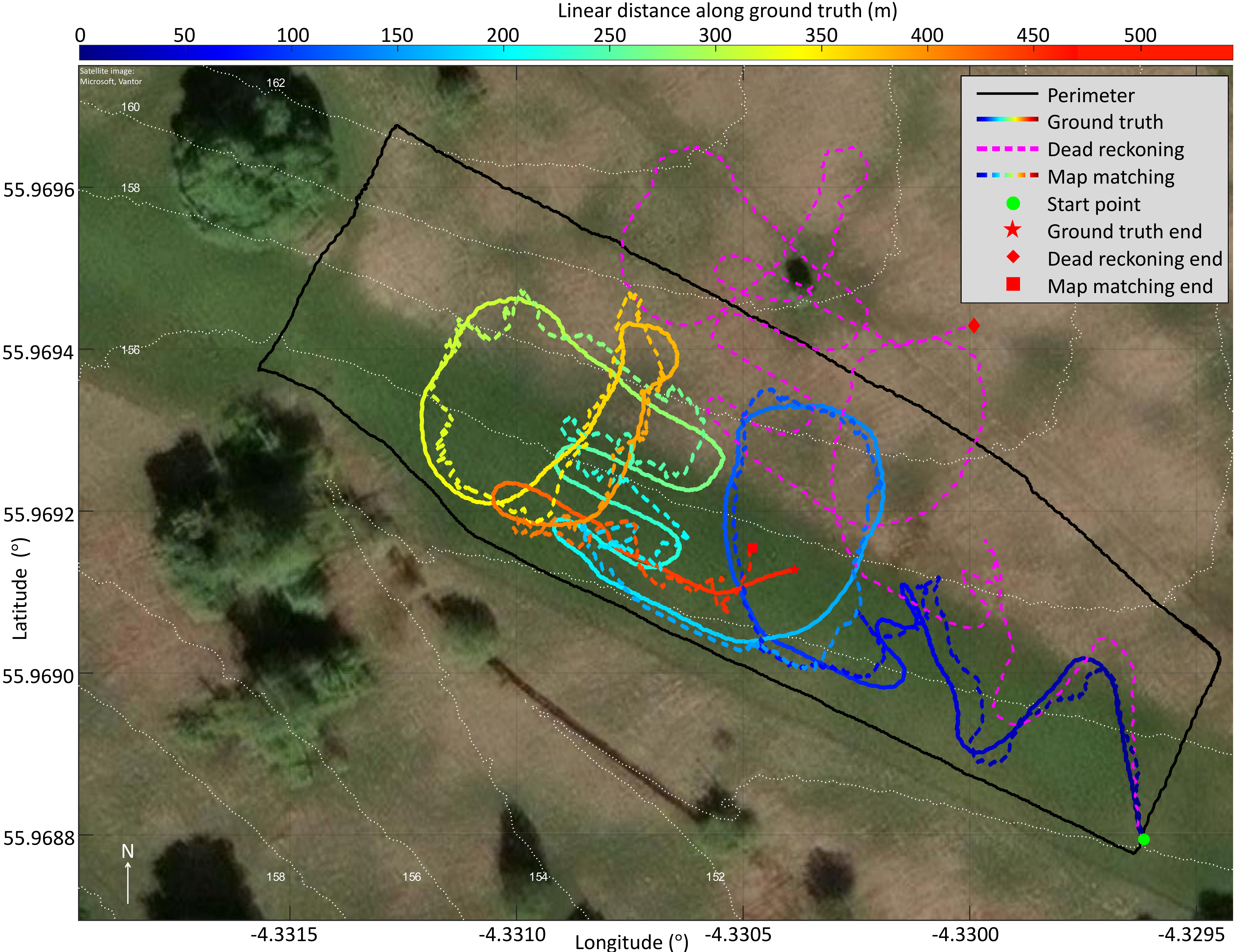}
  \caption{The map matched path. The black line shows the perimeter of the surveyed area. The solid multicoloured line plots the GNSS derived ground truth which begins at the green dot shown in the lower right and ends at the red star. The dashed magenta line plots the path based on dead reckoning and ends at the red diamond. The dashed multicoloured line plots the path predicted by the particle filter based magnetic map matching and ends at the red square.}
  \label{fig:nav_path}
\end{center}
\end{figure*}\\
Figure~\ref{fig:nav_path} shows the test path used to validate the positioning enhancement. The path began at the green dot in the South-East of the test area. The test path consisted of a pseudo-random walk covering a large area of the surveyed area and included doubling back, closed loops, and serpentine segments. The GNSS-based ground truth is shown as the solid multicoloured line terminating at the red square. The evolving line colour indicates the distance travelled along the path. The dashed magenta line ending at the red diamond shows the path predicted by dead reckoning, i.e., a compass heading and a fixed speed. For discussion of this dead reckoning method see Section~\ref{subsec:Particle filter implementation}. The dead reckoning path is similar to the ground truth however, noise in the heading estimate and the variable speed of the true path results in distortions that accumulate over time. The dashed multicoloured line shows the position estimates when magnetic map matching is combined with dead reckoning. The map matched line closely follows the ground truth throughout the path as shown by both, the spatial co-location of the two paths, and the similarity in colour progression along each path.\\    
\begin{figure}[t!]
    \centering
    \begin{tabular}{c}
        \includegraphics[width=0.95\linewidth]{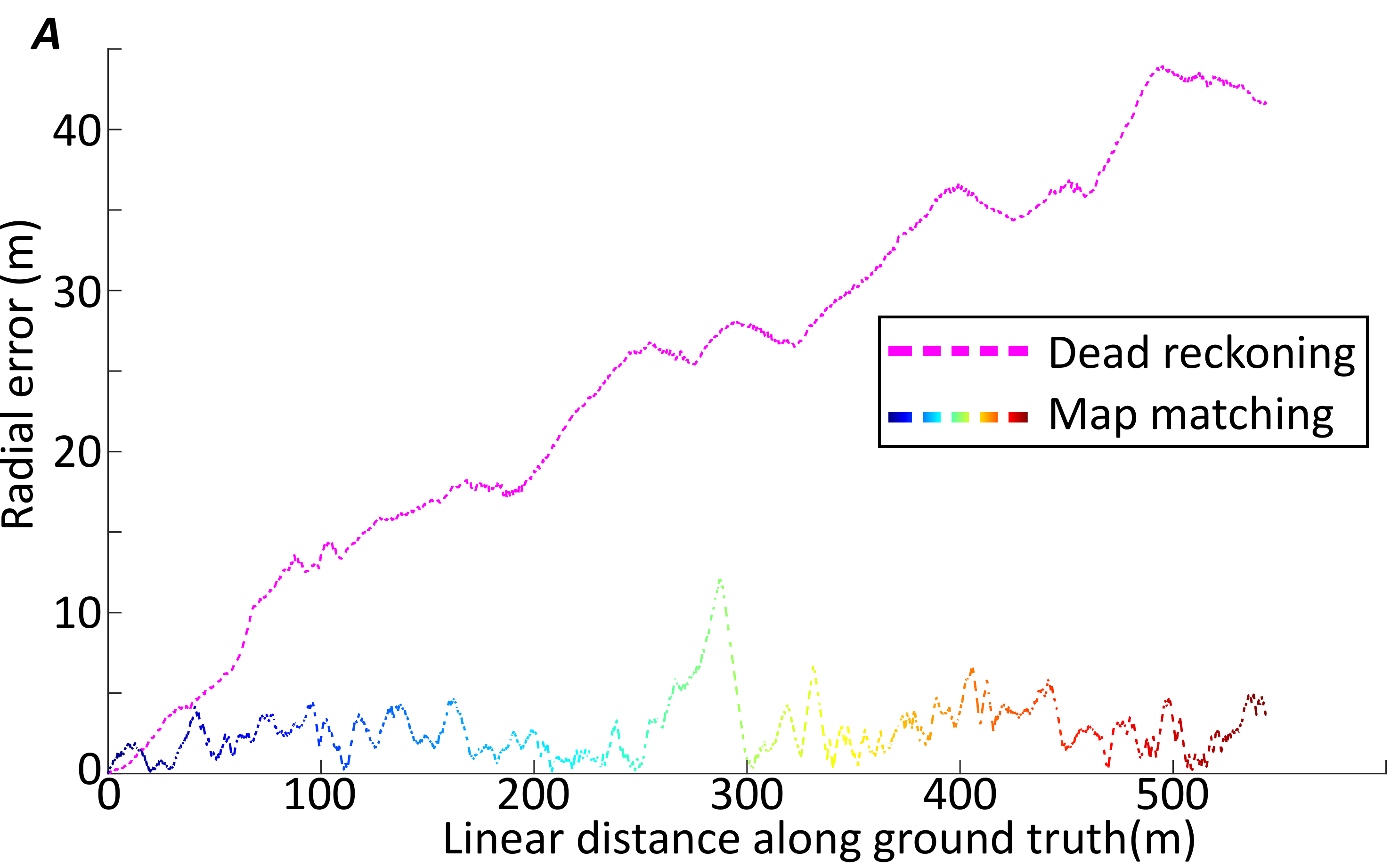}\\
        \cmidrule(lr){1-1}
        \includegraphics[width=0.95\linewidth]{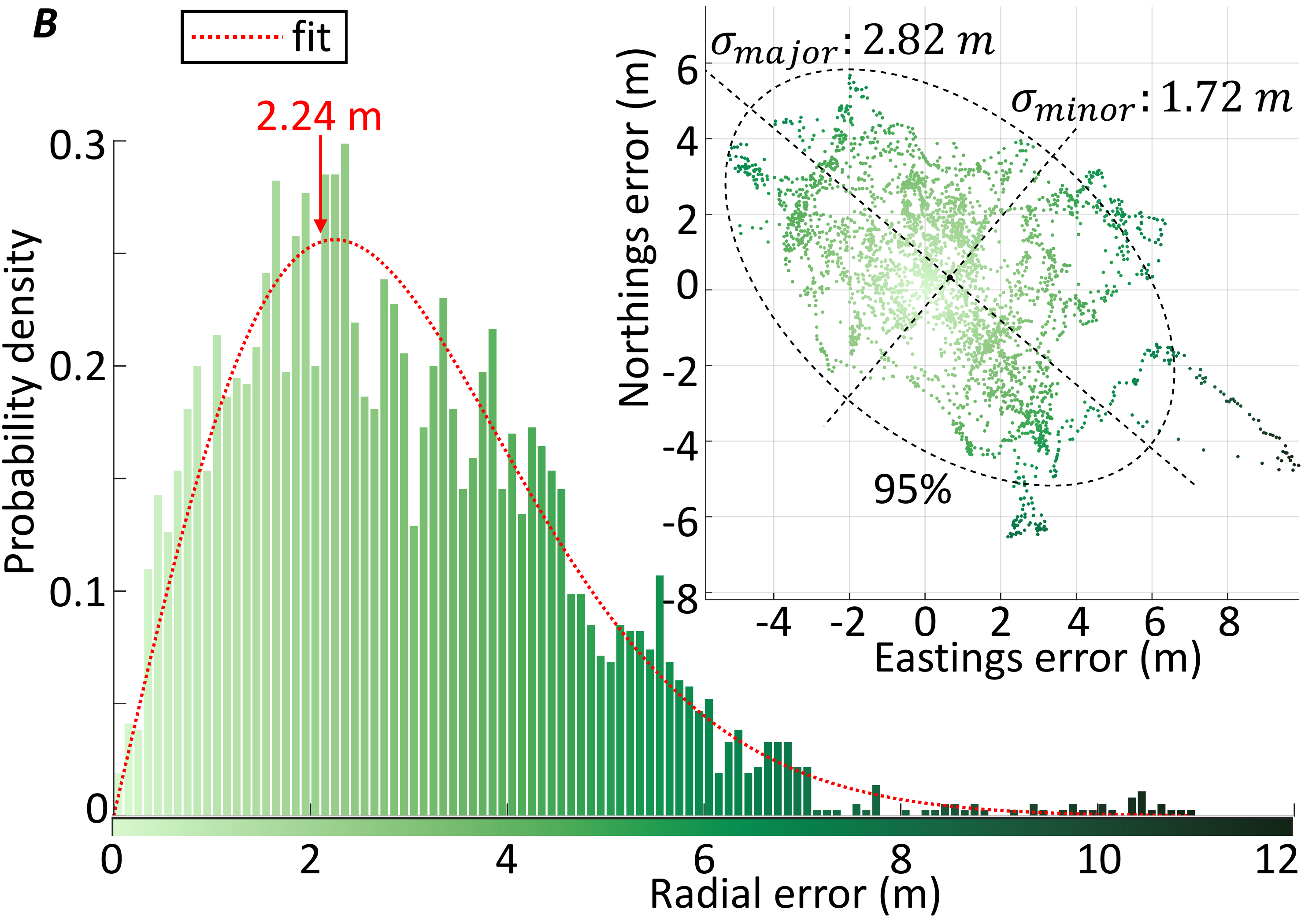}\\
    \end{tabular}
    
    \captionof{figure}{\textit{A} Radial error along the travelled path. The dashed-magenta line corresponds to the dead-reckoning path while the lower dashed-multicoloured line corresponds to the map matching path, mirroring the legend in Fig.~\ref{fig:nav_path}. \textit{B} The main panel shows the normalised distribution of radial errors for the map matching path. The red-dotted line is the fitted Beckmann distribution with a peak probability at 2.24 m. The inset shows a scatter plot of the Eastings and Northings error for each point in the map matched path. The colour-scheme for the points in the inset matches the colour-bar in the main panel. The dotted black lines in the inset show the major and minor axes of the 95th percentile ellipse. Additional error metrics are given in supplementary Table~1.}
    \label{fig: map_match_err}
\end{figure}\\
We quantify the positioning enhancement provided by the map matching using a Beckmann distribution, modelling the positional error vector in the presence of orthogonal errors. Figure~\ref{fig: map_match_err}\textit{A} compares the radial error  between the dead reckoning and map matched paths over the travelled distance. The lines show that the dead reckoning path has a stochastic unbounded error. The addition of the magnetic map matching bounds the error to a measured maximum value of $\sim$12 m. Figure~\ref{fig: map_match_err}\textit{B} shows the distribution of the radial error for the map matched path. The inset in Fig.~\ref{fig: map_match_err}\textit{B} shows the radial error decomposed into its Northings and Eastings components from which the parameters of the Beckmann fit are determined. The orientation of the major and minor axes match the overall orientation of the surveyed area. The dashed red line in the main panel shows the Beckmann fit which closely matches the underlying distribution of radial errors. The peak of the fitted error is 2.24 m, indicating that our OPM based map matching approach is competitive with consumer level single band GNSS positioning systems~\cite{zhong2025quality}. Additionally, the map matching path had a mean radial error of 2.8 m, an endpoint error of 6.5 m, compared with a 41.6 m dead reckoning endpoint error, and a cumulative distribution function 50th percentile error of 2.5 m. Further, analyses from multiple hot start locations and reversed measurement order produced an average Beckmann peak of 2.2 m, see Section~5 of the supplementary for details.   

\section{Methods}\label{sec11}

\subsection{Operating principle of the FID-OPM and system design}
\label{subsec:Operating principle of the FID-OPM and system design}
The OPM is operated in a FID configuration~\cite{hunter2018free, hunter2023optical}. A model of the FID-OPM sensor head is shown in Fig.~\ref{fig:setup}\textit{A}. The core sensing element consists of a silicon MEMS vapour cell containing caesium (Cs) atoms and a nitrogen buffer gas~\cite{dyer2023nitrogen}, which enhances the atomic spin polarization lifetime and, consequently, the achievable sensitivity~\cite{mcwilliam2024optimizing}. The vapour cell is fabricated by etching a silicon wafer using a water-jet process~\cite{dyer2022micro}, producing a cylindrical cavity with a diameter and thickness of 6~mm. The cell is heated using flexible Kapton printed circuit board (PCB) coils, which also generate a polarizing magnetic field, $B_1$, to enhance atomic spin polarization and thereby improve device sensitivity~\cite{hunter2023optical}.\\
\\
The optical system employs two vertical-cavity surface-emitting lasers (VCSELs) with wavelengths near 852~nm. The central VCSEL is current-modulated with a pulse at 500~Hz, switching on and off resonance with the $D_2$ atomic transition to transfer a large population of atoms into the $F = 3$ hyperfine ground state~\cite{hunter2023optical}. A quarter-wave plate converts this light into circular polarization, enabling efficient optical pumping into a highly polarized and sensitive state. The second laser is frequency detuned from resonance and remains linearly polarized, serving as a probe to read out the atomic spin state when the pump beam is off resonance. This light is reflected off a dielectric mirror that is placed directly behind the vapour cell and heater assembly. During readout, the spins precess at the Larmor frequency, which is proportional to the magnitude of the external magnetic field, $B_0$. This precession is detected via optical rotation of the probe beam using a polarimetry system comprising a polarizing beam splitter and a balanced pair of photodiodes.\\
\\
All control signals, including laser and heater modulation and temperature stabilization of the VCSELs via thermoelectric coolers (TECs), are provided by a Digilent Analog-Discovery (AD) 3. The AD3 is also used to digitise the precession signal, which takes the form of a damped sinusoidal oscillation~\cite{hunter2025accurate}. The acquired signal is processed using a software-based processing pipeline that employs a Hilbert transform to extract the precession frequency with high efficiency, enabling magnetic field sampling up to 500~S/s~\cite{hunter2026high}.\\
\\
As shown in Fig.~\ref{fig:setup}\textit{A} the system consisted of four sensors mounted onto a MOLLE-PALS utility harness. The sensors are rigidly mounted on a 1-m, long 15-mm diameter carbon fibre pole for alignment, resulting in a strap-down configuration. The supporting electronics are carried in a chest bag to maximise stand off and ergonomics. The sensors are an RTK enabled GNSS (ArduSimple RTK 2B pro) for ground truth positioning, an Original Equipment Manufacturer (OEM) 3 axis fluxgate magnetometer (Bartington 901), an IMU (adafruit LSM9DS1), and the FID OPM. All systems were powered by two mobile phone powerbanks in parallel, total capacity 20,000~mAh. The full trial consumed $<$~1/2 of the available battery capacity. Additional information on the control and data logging electronics are available in Section~1 of the supplementary material.\\
\\
In addition to the sensors shown in Fig.~\ref{fig:setup}, a base-station was also used, as shown in Section~1 of the supplementary materials. The base station hosted a fluxgate identical to that shown in Fig.~\ref{fig:setup}\textit{A} together with the RTK GNSS base antenna. The base-station fluxgate was used together with the data from Eskdalemuir to monitor space weather effects throughout the trial. Prior to data collection the RTK base antenna was allowed to survey in its position to an accuracy of 1.5 m. The base station was placed well away from the demarcated survey region so that it would not impact the magnetic mapping. 
\subsection{Data processing, calibration, and mapping}
\label{subsec:Data processing, calibration, and mapping}
The data processing chain is: sensor alignment and resampling; application of calibration corrections; data filtering. First, the output from all vector sensors is aligned to a common North East Down (NED) coordinate system. To account for any irregularities in sampling rate, each sensor is independently resampled to its uniform time base using a piecewise cubic Hermite interpolating polynomial. The median sensor sampling rates are 10 Hz, 41 Hz, 248 Hz, and 250 Hz, for the GNSS system, IMU, fluxgate, and OPM respectively.\\
\\
Second, a sensor calibration routine was performed prior to data collection. The primary source of calibration error for the system is the hard-iron and soft-iron error created by the sensors supporting electronics. Specifically, hard-iron error is the offset created by permanently magnetised components of a platform, while soft-iron error is the distortion of Earth's field lines due to the presence of magnetically permeable materials~\cite{creak1882compass}. To address this, on site, but outside of the demarcated survey area, a full-system magnetometer calibration was performed. This consisted of rotation in place several times through 360$^\circ$ clockwise and anti-clockwise, then repetition of this with the torso bent at various angles - gathering samples with the sensors in the largest range of orientations practical. To correct for the hard-iron and soft-iron errors, a 3D ellipsoid was fitted to the aligned and resampled calibration data from the fluxgate (see supplementary Section~2). The centre position of the ellipsoid fit corresponds to the hard-iron offset correction with the soft-iron correction matrix derived from the radii of the ellipse. Further, the impact of the platform errors on the OPM is suppressed using the Tolles-Lawson algorithm~\cite{jukic2024applications}. Specifically, the Earth's field at sample $i$ ($B_{0}^{i}$) is given by,
\begin{equation}
    B_{0}^{i} = B_{\text{opm}}^{i}-A^{i}x,
    \label{eqn:Tolles_Lawson}
\end{equation}
where $B_{\text{opm}}^{i}$ is the ith OPM value, $A^{i}$ is an 18 element vector related to the platform's orientation, and $x$ is the 18 element Tolles-Lawson coefficient vector. First, the corrected fluxgate data from the calibration routine is used together with the resampled OPM data gathered at the same time to derive the coefficients of $x$ using least squares fitting. $A^{i}$ is,
\begin{equation}
\begin{aligned}
  A^{i} = &[\cos(\theta^i),\cos(\phi^i),\cos(\psi^i)\\
    &B_{\text{opm}}^{i}\cos^2(\theta^i),B_{\text{opm}}^{i}\cos^2(\phi^i),B_{\text{opm}}^{i}\cos^2(\psi^i),\\&B_{\text{opm}}^{i}\cos(\theta^i)\cos(\phi^i),B_{\text{opm}}^{i}\cos(\theta^i)\cos(\psi^i),\\
    &B_{\text{opm}}^{i}\cos(\phi^i)\cos(\psi^i),\dot{B}_{\text{opm}}^{i}\dot{\cos}(\theta^i)\cos(\theta^i),\\
    &\dot{B}_{\text{opm}}^{i}\dot{\cos}(\phi^i)\cos(\theta^i),\dot{B}_{\text{opm}}^{i}\dot{\cos}(\psi^i)\cos(\theta^i),\\
    &\dot{B}_{\text{opm}}^{i}\dot{\cos}(\phi^i)\cos(\phi^i),\dot{B}_{\text{opm}}^{i}\dot{\cos}(\theta^i)\cos(\phi^i),\\
    &\dot{B}_{\text{opm}}^{i}\dot{\cos}(\psi^i)\cos(\phi^i),\dot{B}_{\text{opm}}^{i}\dot{\cos}(\phi^i)\cos(\psi^i),\\
    &\dot{B}_{\text{opm}}^{i}\dot{\cos}(\theta^i)\cos(\psi^i),\dot{B}_{\text{opm}}^{i}\dot{\cos}(\psi^i)\cos(\psi^i)].  
\end{aligned}    
\label{eqn:A_Matt}
\end{equation}
The dot-notation represents the time derivative. The directional cosine terms are derived from the corrected fluxgate measurements as each component over the magnitude of the field,
\begin{equation}
\begin{aligned}
  \cos(\theta^i),\cos(\phi^i),\cos(\psi^i) &=\frac{B^{i}_{N,E,D}}{|B^{i}_{NED}|}
\end{aligned}    
\label{eqn:d_cos}
\end{equation}
In addition to the magnetic calibration, the accelerometers and gyroscopes in the IMU were calibrated by having the individual stand upright and stationary. The mean IMU sensor offsets were then subtracted from all subsequent data.\\
\\
Third, the mapping and test path data was gathered. As shown in Fig.~\ref{fig:survey_map_no_interp} an area of 160 m~$\times$~40 m was surveyed. Non-magnetic markers were used to lay out a grid of survey lines in 10 m intervals. In addition to the raster grid, two diagonal algorithm calibration lines were walked, the use of which is discussed in Section~\ref{subsec:Particle filter implementation}. After the mapping pass, the test path shown in Figs.~\ref{fig:setup} and~\ref{fig:nav_path} was walked.\\ 
\\
The data from the mapping and test paths were passed through the aforementioned alignment and resampling process. The calibration factors were then applied to produce corrected IMU, fluxgate, and OPM measurements at their respective median sample rates. The fluxgate and OPM data were lowpass filtered with a cutoff of 20~Hz to remove any high-frequency, anthropogenic noise sources in the data. The IMU, fluxgate and OPM data were downsampled to 10~Hz using a non-overlapping mean filter so that all data were co-registered with the recorded GNSS positions. Finally, a reference magnetic field value of 50208.8~nT was subtracted from the OPM data to convert the measured fields into relative anomalies. The OPM data from the mapping pass, excluding the algorithm calibration lines, were used to construct an interpolation function $F$(Lat, Lon). The function takes in latitude and longitude coordinates and returns an interpolated magnetic field value.  
\subsection{Particle filter implementation}
\label{subsec:Particle filter implementation}
We realise our magnetic map matching based position estimation enhancement through the use of a particle filter~\cite{jukic2024quantum}. The particle filter operates in the steps given in Algorithm~\ref{pf_algo}. For additional details please see Sections~3 and 4 of the supplementary.
\begin{algorithm}
\caption{Particle filter implementation}\label{pf_algo}
\begin{algorithmic}[1]
\State Using dead reckoning, move from the initial estimated position $P_i$ to a new estimated position $P_{i+1}$.
\[
P_{i} \xrightarrow[]{\text{Dead reckoning}} P_{i+1}.
\]
\State Spawn a swarm of particles $S$ around $P_{i+1}$ according to a statistical distribution and a process noise parameter.
\State Generate a magnetic sample at the position of each particle using the interpolating map function $F(\mathrm{Lat}, \mathrm{Lon})$.
\State Compare each particle sample with the magnetic value measured by the OPM.
\State Assign each particle a weight according to the measurement likelihood.
\State Compute the weighted sum of the particles to update the estimated position:
\[
P_{i+1}\xrightarrow[]{\text{Map matching}} P_{i+1}^{\prime}.
\]
\State Set $P_i \Leftarrow P_{i+1}^{\prime}$ and repeat the procedure for the next iteration.
\end{algorithmic}
\end{algorithm}\\
\\
Our dead reckoning method is based on a compass bearing and a constant speed. We recognise that this is not representative of all situations. However, movement along a compass bearing is a reasonable approximation for the majority of platforms that move on surfaces. Further, we use a constant speed rather than an integrating speed estimator due to the performance limitations of our IMU. The accelerometer bias in the IMU results in a speed based path which diverges rapidly from the true path. This divergent path is then both a poor input for the particle filter and an unrealistic point of comparison for Fig.~\ref{fig:nav_path}. For the dead reckoning we use a constant speed of 1.4 m.s$^{-1}$. This is derived from the median GNSS speed during the survey phase only. The standard deviation in measured speed during the survey phase is $0.4$ m.s$^{-1}$. As reference, the median speed from the GNSS ground truth during the test path was 1.5 m.s$^{-1}$ with a standard deviation of $0.3$ m.s$^{-1}$.\\  
\\
The result of the dead reckoning path is shown by the magenta dotted line in Fig.~\ref{fig:nav_path}. Specifically, we fuse the fluxgate magnetometer and accelerometers from the IMU to create a compass. For each iteration of the algorithm, $P_{i+1}$ is derived from $P_{i}$ by taking a step of fixed length in the direction given by the compass. Using the IMU to only correct for the orientation of the fluxgate (creating a compass) does not require the integration of the accelerometer data over time, reducing the impact of the bias error.\\
\\
Correct setting of the process noise and measurement likelihood are important for the performance of the particle filter. To set these parameters, we use the data gathered in the diagonal cross lines mentioned in Fig.~\ref{fig:survey_map_no_interp} and Section~\ref{subsec:Data processing, calibration, and mapping}. Specifically, analogous to the sensor calibration step described in Section~\ref{subsec:Data processing, calibration, and mapping}, we use the algorithm calibration lines to gather data that is used to optimise the parameters of the particle filter.\\
\\
The spatial distribution of the particle swarm around $P_{i+1}$ is set by the process noise and the weights from the previous iteration. Specifically, each swarm (for $i>$ 0) is resampled from the previous swarm where particles with higher weights are more likely to be selected multiple times, while lower weight particles fall away. This allows the spatial distribution of the swarm to dynamically evolve around the highest likelihood points as the path progresses. To ensure that the swarm retains a sufficient degree of spatial diversity, the position of each particle is independently perturbed at each iteration by a random number sampled from a Normal distribution with $\mu_1 = 0$ and $\sigma_1$ = the process noise. Too small a process noise results in a swarm that does not sample a sufficiently large region around $P_{i+1}$, while too large a swarm can be misled by sampling distant regions of the map. To set the process noise, we use the algorithm calibration paths by converting the associated GNSS ground-truth and corresponding dead reckoning into two sequential lists of $\Delta$ latitude, $\Delta$ longitude, i.e, the change in coordinate per step for each path. By subtracting the dead reckoning based list from the GNSS based list we find the single largest misstep in the dead reckoning path during the algorithm calibration. Setting the process noise to this value means that the particle swarm always retains sufficient spatial diversity to absorb the largest observed misstep. Here, the process noise is $\sigma_1 = 3.48\times10^{-6~\circ}$ ($\approx 0.3$ m radial error at the latitude of the test site).\\
\\
To determine the measurement likelihood function, we compare the measured OPM data from the algorithm calibration lines to the values generated by the interpolating map function $F$ at the corresponding GNSS ground-truth coordinates. This process effectively quantifies the interpolation discrepancy between the measured OPM values and the magnetic map. From our calibration, we found that the differences were Normally distributed with $\sigma_2 = 36$ nT, making the navigation map-limited.\\ 
\\
At the start of the map matching path, a `hot-start' \cite{paonni2010performance} is assumed, in which $P_0$ coincides with the GNSS ground truth at the start of the trial. This assumption is consistent with many real world scenarios in which GNSS is only denied/degraded for a portion of the travelled path. Outside of the initial mapping stage, the algorithm calibration, and the hot-start, the GNSS ground truth is not used by the particle filter in any way. The GNSS ground truth is only included in Fig.~\ref{fig:nav_path} for visual comparison. \\ 
\\
The particle filter uses 1500 particles, and a particle resampling ratio of 0.5. Here, the particle filter is implemented in post processing. However, the $\sim$6 minutes of test path data (3648 samples) is processed in $<1$ minute using an Intel Ultra 7 155H (laptop) CPU, making the approach compatible with real-time implementations.

\section{Discussion}\label{sec12}
These results directly quantify the advantage in GNSS-independent positioning accrued when this wearable free-induction optically pumped magnetometer is added to the conventional MEMS inertial sensors used for dead-reckoning position. Comparison of the radial error accrued with and without the OPM, shown in Fig.~\ref{fig: map_match_err}, reveals that the addition of the OPM data to constrain the particle filter alters the distribution of errors with time. Without the benefit of OPM data, divergent radial errors are obtained, whereas with the inclusion of OPM data, these errors are bounded, converging on a Beckmann distributed error ellipse with major and minor mean errors of 2.82~m and 1.72~m respectively. The orientation of major and minor error ellipse axes broadly follows the directions of lower and higher contrast in the observed crustal magnetic anomaly map (Fig.~\ref{fig:survey_map_no_interp}).\\ 
\\
This work directly set out to examine a key question in the application of quantum technologies to enhanced positioning; can magnetic map matching improve positioning bounds even in the presence of anthropogenic magnetic fields, and what scale of spurious or malicious magnetic signals would need to be generated to degrade the value of this positioning constraint? For this reason an essential feature of the experiment was that the magnetic anomalies referenced should be naturally occurring, and extend over a significant area. Anomalies on the scale of those recorded (hundreds of nT at ground level) are widespread across the Earth's surface, and although man-made permanent and generated magnetic dipoles can easily exceed this field magnitude, to do so over distances of tens or hundreds of square metres requires infeasibly large quantities of magnetic material or electrical energy. The man-made magnetic feature at 55.9692~N -4.3300~W, generated an anomaly of significant magnitude, but of small spatial extent and did not dominate the results of the positioning analysis.\\
\\
The measurement of positioning constraint relative to natural crustal anomalies is therefore an essential aspect of this study, with consequential constraints on the experimental hardware and conditions available. The presented OPM sensitivity and accuracy (Fig.~\ref{fig:a_dev}) is that of the wearable, battery powered sensor that was actually used, rather than laboratory test results for the same sensor type~\cite{hunter2023optical}. Likewise, our positioning analysis is not a direct comparison of inertial dead-reckoning to magnetic map matching, but rather a comparative measurement of positioning error accrued by dead-reckoning with and without additional OPM data. Design of a system for inertial dead-reckoning without size, weight, power and cost constraints may accomplish superlative performance, particularly where emergent quantum sensing and timing systems are used~\cite{li2026,Li2025}. Further, our methodology is not intended as an outright replacement for GNSS under all circumstances. Rather, our use of GNSS is selective, employing it for the effective creation of magnetic maps and algorithm calibrations such that meaningful position estimation augmentation can be made in its absence. This study should be read as a measurement of additional positioning benefit when a stable OPM sensor is added to a wearable positioning system, with traceability to the design and performance of the sensor system. This rationale also underpins the choice of particle filtering algorithm selected, although we note a number of advanced alternative approaches~\cite{Sengupta2025}.\\    
\\
These results rely on effective mitigation of platform noise sources. As longer-term measurement drift most detrimentally impacts positioning by map matching, the platform's hard- and soft-iron errors were most thoroughly suppressed (Section \ref{subsec:Data processing, calibration, and mapping}, and Supplemental Material), making use of the low-drift OPM sensor feasible. Generalised use of an OPM in this manner in platforms of various types can be carried out, with a variety of approaches for extension of Tolles-Lawson algorithms recently published~\cite{jukic2024applications,Du2019,Zhai2023}.  

\section{Conclusion}\label{sec_conclusion}

Augmentation of positioning in the absence of GNSS using magnetic map matching covers a wide range of platforms, use cases and technical approaches, and any new capability in this area will rest on successful integration of sensing, platform noise mitigation, magnetic map database, interpolation and positioning algorithm with an extant navigation system. Any practical validation of positioning enhancement, such as this one, will constitute an instantiation of experimental choices made regarding platform, environment, form factor, duration and other factors. However, the results presented, showing that a stable OPM magnetometer may, with appropriate mitigation of platform effects and prior availability of map data, add a measurable enhancement to dead-reckoning positioning accuracy, fundamentally changing the distribution and magnitude of radial errors, and that this may be carried out with reference to permanent, natural crustal geomagnetic anomalies. Elements of these approaches may be generalised to the wider application of GNSS-resilient positioning around the globe.

\backmatter

\bmhead{Supplementary information}
This manuscript is accompanied by 1 supplementary document.



\bmhead{Acknowledgements}
The authors would like to thank Mugdock Country Park, East Dumbartonshire Council, Scotland, for land access.\\
\\
The results presented in this paper rely on the data collected
at Eskdalemuir observatory. We thank British Geological
Survey for supporting its operation and INTERMAGNET
for promoting high standards of magnetic observatory practice
(www.intermagnet.org).



\section*{Declarations}


\subsection*{Funding}
This work was partly funded by the EPSRC UK Quantum Technology Hub in Quantum Enabled Position, Navigation and Timing (QEPNT, EP/Z533178/1).
\subsection*{Conflict of interest}
The authors declare no conflicts of interest.
\subsection*{Data availability}
The datasets generated during and/or analysed during the current study are available from [https://doi.org/10.15129/b662fc69-774e-4235-b450-c4e1d2a870df].
\subsection*{Code availability}
The code generated during and/or used during the current study are available from the corresponding author on reasonable request.
\subsection*{Author contribution}
S.S. and S.I. conceived the experiment. A.M., D.B., J.M. and P.G. developed and fabricated the MEMS cell of the FID OPM. S.S. D.H. C.D. M.M. S.I. and E.R. developed the FID OPM and experimental system. S.S., D.H. and S.I. performed the field trial with input from P.U. S.S. performed the data analysis. S.S., D.H., C.D., M.M., P.U., P.G., J.M., E.R. and S.I. contributed to manuscript writing, reviewing, and editing.

\bibliography{sn-bibliography}

\end{document}